\documentclass[12pt,oneside]{article}
\usepackage[a4paper,centering,left=2.5cm,top=3cm]{geometry}
\usepackage[utf8]{inputenc}
\usepackage{graphicx}
\usepackage{color}
\usepackage{subscript}
\usepackage{amsmath,amssymb}
\usepackage{authblk}

\begin{document}

\title{The role of ion transport phenomena in memristive double barrier devices}

\author[1]{Sven Dirkmann\thanks{Corresponding author: sven.dirkmann@rub.de}}\author[2]{Mirko Hansen}\author[2]{Martin Ziegler}\author[2]{Hermann Kohlstedt}\author[1]{Thomas Mussenbrock}

\affil[1]{Lehrstuhl für Theoretische Elektrotechnik, Fakult\"at f\"ur Elektrotechnik und Informationstechnik, Ruhr-Universit\"at Bochum, D-44780 Bochum, Germany}\affil[2]{Nanoelektronik, Technische Fakult\"at, Christian-Albrechts-Universit\"at zu Kiel, D-24143 Kiel, Germany}

\date{\normalsize\today}

\maketitle

\begin{abstract}
In this work we report on the role of ion transport for the dynamic behavior of a double barrier quantum mechanical Al/Al$_2$O$_3$/Nb$_{\text{x}}$O$_{\text{y}}$/Au memristive device based on numerical simulations in conjunction with experimental measurements. The device consists of an ultra-thin Nb$_{\text{x}}$O$_{\text{y}}$ solid state electrolyte between an Al$_2$O$_3$ tunnel barrier and a semiconductor metal interface at an Au electrode. It is shown that the device provides a number of interesting features for potential applications such as an intrinsic current compliance, a relatively long retention time, and no need for an initialization step. Therefore, it is particularly attractive for applications in highly dense random access memories or neuromorphic mixed signal circuits. However, the underlying physical mechanisms of the resistive switching are still not completely understood yet. To investigate the interplay between the current transport mechanisms and the inner atomistic device structure a lumped element circuit model is consistently coupled with 3D kinetic Monte Carlo model for the ion transport. The simulation results indicate that the drift of charged point defects within the Nb$_{\text{x}}$O$_{\text{y}}$ is the key factor for the resistive switching behavior. It is shown in detail that the diffusion of oxygen modifies the local electronic interface states resulting in a change of the interface properties of the double barrier device.  
\end{abstract}

\newpage

\section{Introduction}

The research in the field of memristive devices dates back to the 1970s when Chua introduced his theory of memristors and memristive devices.\cite{Chua:1971} This idea has emerged a considerable amount of interest after 2008 when Strukov et al. linked their resistive switching device to Chuas's theory.\cite{Strukov:2008} Today, memristive (or synonymously resistive switching) devices have been identified as promising candidates for future non-volatile memory applications due to their distinct key features which are i) low power consumption, ii) passivity, and iii) scalability into the nanometer scale.\cite{Yang:2013} Beyond their applications as non-volatile memories, resistive switching devices turned out to be applicable as artificial synapses in neuromorphic circuits.\cite{Ziegler:2012, Thomas:2013, Zahari:2015}

Many of the available metal-oxide resistive switching devices rely on the stochastic phenomenon of creation and rupture of conductive metal filaments within a solid state electrolyte matrix. Inherent are initial electroforming procedures and distinct ``on'' and ``off'' states. The latter are in fact beneficial for memory applications. However, in the context of neuromorphic applications a wider dynamic range is required, rather than a purely ``digital'' behavior. To overcome these limitations various kinds of interface-based resistive switching devices have been developed. Predominantly the switching mechanisms rely either on the change of height of a tunnel barrier, and thus on the electron tunneling probability, or on the change of a Schottky contact.\cite{Baik:2010, Meyer:2008, Fujii:2005, Wang:2013} Recently Hansen et al. presented a memristive device consisting of both, a tunnel barrier and a Schottky contact which embed a thin solid state electrolyte.\cite{Hansen:2015} This device provide specific features such as a highly uniform current distribution, an intrinsic current compliance, an improved retention time, and no need for an initial electroforming procedure. These features make the device interesting for its application as artificial synapses in bio-inspired neural networks. However, the physical mechanisms which are responsible for the memristive behavior are still not completely understood. To explain the physics two hypotheses are proposed: The first assumes charging and de-charging of trap states within the solid state electrolyte and/or at the metal-semiconductor interface to be the main reason for the resistance change.\cite{Simmons:1978, Odagawa:2004, Shao:2015} The second hypothesis is that the motion of charged point defects within a sufficiently high applied electric field controls the properties of the tunnel barrier, and therefore the resistance of the device.\cite{Seong:2008, Sawa:2008}

The aim of this work is to provide a model which allows to investigate the role of the ion transport for the resistive switching of the double barrier memristive device. We utilize a 3D kinetic Monte Carlo code in order to describe the ion transport within the solid state electrolyte subject to both externally applied and Coulomb fields.\cite{Dirkmann:2015} The kinetic Monte Carlo model is consistently incorporated into an efficient lumped element circuit model of the device. It is important to note that the simulation parameters are chosen in accordance with an experimental setup. We find that the ion transport is a key factor for the resistive switching of the device. Furthermore, we identify an adsorption mechanism of ions at the Au electrode to be a key factor for the retention characteristics of the device. The comparison of the simulation results with experimental findings shows excellent agreement.

\section{Simulation approach}

\begin{figure}[t!]
\centering
\includegraphics[width=12cm]{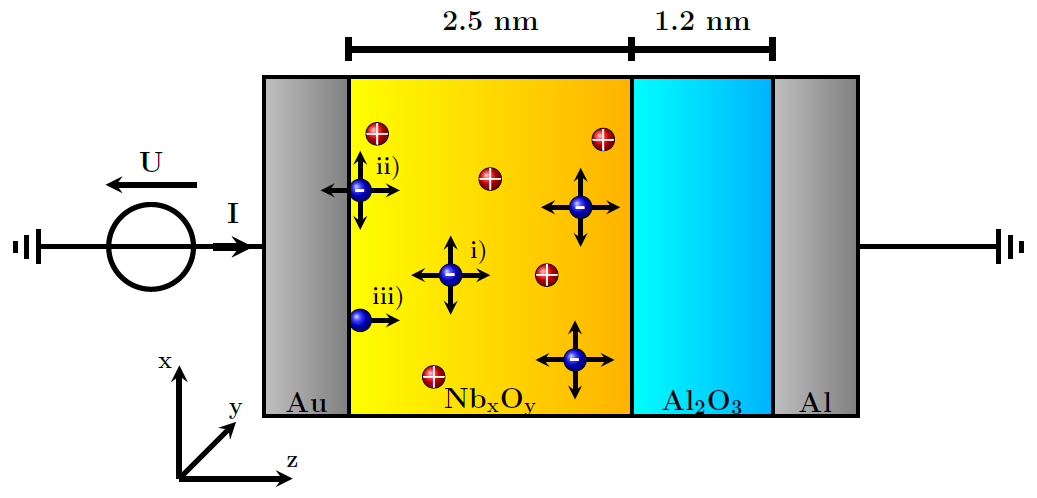}
\caption{Simulation domain of the double barrier memristive device, consisting of an Al$_2$O$_3$ tunnel barrier, a Nb$_{\text{x}}$O$_{\text{y}}$ solid state electrolyte, and a Nb$_{\text{x}}$O$_{\text{y}}$/Au Schottky contact. Positively charged (red circles) and negatively charged (blue circles) defects are located within the solid state electrolyte. Positively charged defects are assumed to be immobile, negatively charged defects are allowed to move within the solid state electrolyte (i), adsorb at the Au electrode (ii), and desorb from the Au electrode (iii).
\label{MemristiveDevice}}
\end{figure}

The simulation scenario is shown in figure \ref{MemristiveDevice}. It consists of a memristive device with the layer sequence Au/Nb$_{\text{x}}$O$_{\text{y}}$/Al$_2$O$_3$/Al. We assume the Nb$_{\text{x}}$O$_{\text{y}}$ layer to act as an ionic/electronic mixed conductor. It represents a solid state electrolyte in which mobile oxygen ions (blue circles) drift under the influence of an externally applied electrical field. In order to guarantee for charge neutrality of the simulation domain, the negative oxygen ions are compensated by stationary positive ions (red circles).\cite{Kittel:2004} At the initial simulation stage the ions are randomly distributed within the Nb$_{\text{x}}$O$_{\text{y}}$ assuming thermodynamic equilibrium. The Nb$_{\text{x}}$O$_{\text{y}}$/Au interface is assumed to be inert to prevent oxidation. It can be regarded as a Schottky contact. The Al$_2$O$_3$ tunnel barriers are electrically high-quality barriers enabling elastic electron tunneling. They are nearly free of defects and highly stable. Both the tunnel barrier and the Au/Nb$_{\text{x}}$O$_{\text{y}}$ interface represent chemical barriers which confine the ions within the Nb$_{\text{x}}$O$_{\text{y}}$ layer. Details about the devices fabrication can be found in Hansen et al. \cite{Hansen:2015}.
 
Before we describe the details of the model it might be helpful to firstly explain the overall idea of the simulation approach: The physical device is decomposed into distinct lumped elements which are connected in series, similar to an earlier approach.\cite{Hansen:2015} The tunnel barrier is mimicked by a voltage controlled current source based on the famous Simmons formula \cite{Simmons:1963}. The Nb$_{\text{x}}$O$_{\text{y}}$/Au interface is descibed the Schottky contact model.\cite{Sze:2007, Tung:2014} The solid state electrolyte is modeled by an Ohmic resistance which depends on the inner atomic structure. As stated before, the transport of ions within the solid state electrolyte under the influence of the externally applied electric field and the Coulomb field due to the ions themselves is of particular importance. Therefore, the lumped element circuit model is consistently coupled with a 3D kinetic Monte Carlo (KMC) model for the ion transport in such a way that Kirchhoff's voltage and current laws for the circuit model are satisfied at all instances of time. Within an inner loop the dynamical state of the solid state electrolyte is calculated subject to the electric field acting on the ions and subject to a set of appropriate boundary conditions. From this the ohmic resistance of and the voltage drop across the solid state electrolyte are calculated. Applying Kirchhoff's current law which stated that the current through the device is constant, the individual voltage drops across the tunnel barrier and the Schottky contact as well as the current itself are calculated within an outer loop.  

Now we come to the details of the model: Using X-ray diffraction measurements the Nb$_{\text{x}}$O$_{\text{y}}$ layer has found to be amorphous.\cite{Hansen:2015} Due to the short range periodicity of amorphous material it is reasonable to model this layer by a 3D lattice with a lattice constant corresponding to the hopping distance of oxygen ions in Nb$_{\text{x}}$O$_{\text{y}}$.\cite{Dirkmann:2015, Menzel:2015} We use lattice constants of 0.33 nm in $z$ direction and and 0.25 nm in $x$ and $y$ direction. This is indicated in figure \ref{MemristiveDevice}. It takes into account the dimensional lattice mismatch of Nb$_{\text{x}}$O$_{\text{y}}$. The 3D simulation domain consists of a quadratic 9 nm $\times$ 9 nm basis area, a 2.5 nm thick solid state electrolyte and a 1.2 nm thick tunnel barrier (see figure \ref{MemristiveDevice}). 

\begin{table}[t!]
\centering
\begin{tabular}{l l}
\hline 
\hline
Physical quantity & Value \\
\hline
Temperature & 300 K \\
Phonon frequency & $1.0 \times 10^{12}$ Hz \\
Lattice constant $(x, y)$ (Nb$_{\text{x}}$O$_{\text{y}}$) & $3.3\times 
10^{-10}$ m \\
Lattice constant $(z)$ (Nb$_{\text{x}}$O$_{\text{y}}$) & $2.5\times 10^{-10}$ m 
\\
Relative permittivity (Al$_2$O$_3$) & 9 \\
Relative permittivity (Nb$_{\text{x}}$O$_{\text{y}}$) & 42 \\
Conductivity (Nb$_{\text{x}}$O$_{\text{y}}$) & $2 \times 10^{-4}$ 
$\frac{1}{\Omega \text{m}}$ \\
E$_{ij}$ for ion diffusion & 0.68 eV \\
E$_{ij}$ for desorption & 0.25 eV \\
Defect density n & $5\times 10^{20}$ $\frac{1}{\text{cm}^3}$\\
Tunnel barrier height $\Phi_0$ & 3.1 eV \\
\hline
\hline
\end{tabular}
\caption{Parameters for the simulation}
\label{Table}
\end{table}

For the KMC model which is described in detail in ref. \cite{Fichthorn:2000, Voter:2007, Dirkmann:2015} three different hopping processes of the oxygen ions have been taken into account (see figure \ref{MemristiveDevice}): i) diffusion within the solid state electrolyte, ii) adsorption  and iii) desorption (iii) at the respective interfaces of the Nb$_{\text{x}}$O$_{\text{y}}$ layer. The related hopping rates k$_{ij}$ for the KMC model are given by an Arrhenius law,
\begin{align}
k_{ij} = \nu e^{-\frac{E_{ij}}{k_B T}}.
\end{align}
$\nu$ is the phonon frequency, $T$ is the lattice temperature, $k_B$ is the Boltzmann constant, and $E_{ij}$ is the hopping (potential) barrier between the lattice sites $i$ and $j$. For the hopping barrier we use $E_{ij} = E^k_0 + \frac{1}{2}(E_j - E_i)$, where  $E^k_0$ is the activation energy for the hopping process ($k$ = diffusion, adsorption, desorption). The term $\frac{1}{2}(E_j - E_i)$ represents a linear correction term of the hopping barrier due to an externally applied electric field and the local Coulomb field due to the ions themselves. 

The transport of ions within the solid state electrolyte evolves as follows: After inserting a certain number of stationary positive and mobile negative ions into the Nb$_{\text{x}}$O$_{\text{y}}$ layer, the negative ions drift according to the local electric field which itself is a superposition of the field due to the externally applied voltage and the Coulomb field of the remaining ions in the system. When an oxygen ion reaches one of the interfaces a surface interaction takes place, depending on the local electric potential. The simulation parameters, which are mainly extracted from the experiment, are collected in table \ref{Table}.

The first step of the simulation loop is to calculate the electric potential within the Al$_2$O$_3$ tunnel barrier. We assume that the Al$_2$O$_3$ is a perfect insulator which contains no electric charge carriers. Therefore, the electric potential can be obtained from a simple Laplace equation subject to Dirichlet conditions,
\begin{gather}
\nabla^2 \Phi(\vec{r}) = 0.
\label{Poisson}
\end{gather}
The permittivity of the Al$_2$O$_3$ layer is assumed to be homogeneous. This leads of course to a spatially linear potential and a spatially constant electric field. With the voltage drop across the tunnel barrier the local electron tunneling current density can be calculated from the Simmons formula,
\begin{gather}
J_{Tunnel} = \frac{q}{2 \pi h d_{eff}^{2}} \left(  x_1 \cdot \exp\{ -\frac{4 \pi d_{eff} \sqrt{2m}}{h} \sqrt{x_1} \} -  x_2  \cdot \exp\{-\frac{4 \pi d_{eff} \sqrt{2m}}{h} \sqrt{x_2}\} \right),
\label{eq1}
\end{gather}
with $x_1= q\Phi_0 - \frac{q V_{Tunnel}}{2}$ and $x_2= q\Phi_0 + \frac{q V_{Tunnel}}{2}$. $\Phi_0$ is the barrier height of the tunnel barrier ($\Phi_0$ = $\frac{\Phi_{\text{Al}} + \Phi_{\text{Nb}_{\text{x}}\text{O}_{\text{y}}}}{2} \approx 3.1$ eV). The elementary charge, the free electron mass, and the Planck constant are given by $q$, $m$, and $h$ respectively. $d_{eff}$ is the effective thickness of the tunnel barrier. Since the Simmons formula accounts only for elastic tunneling, the local ion concentration at the Al$_{2}$O$_{3}$ is of particular importance for the resulting electron tunneling current. In order to take this into account the tunnel distance has been implemented as an effective distance $d_{eff}$, which itself depends on the position of the negative ions within the Nb$_{\text{x}}$O$_{\text{y}}$ layer, 
\begin{align}
d_{eff} = d_0 + \lambda_d \cdot \bar{d}.
\label{Tunnellength}
\end{align}
$\lambda_d$ is a fit parameter, $\overline{d}$ is the average relative change of the oxygen ion position from their inertial position, and $d_0$ is the initial effective thickness of the Al$_2$O$_3$ layer ($d_0$ = 1.3 nm). 

The electric potential within the Nb$_{\text{x}}$O$_{\text{y}}$ layer cannot be obtained directly from Poisson's equation as the charge density within the layer is not known a priori. The assumption of an electric charge carrier-free insulator, as used for the tunnel barrier, is not valid in the Nb$_{\text{x}}$O$_{\text{y}}$ layer due to an electric current which is present here. Instead, we calculate the local electric potential by solving the continuity equation $\nabla \cdot \vec{J} = 0$ in conjunction with a constitutive law that couples locally the electric field $\vec{E}$ and the current density $\vec{J}$. The required relation between the electric field and the current density is given by a local Ohm's law, $\vec{J}(\vec{r}) = \sigma\vec{E}(\vec{r})$.  Here $\sigma$ represents the electric conductivity of the Nb$_{\text{x}}$O$_{\text{y}}$ layer, which is assumed to be homogeneous. It is important to note that the displacement current can be neglected because it scales with $\left(L/T c \right)^2$ where $c$ is the speed of light and $L$ and $T$ are the length and the time scales of the system. The displacement current is therefore small compared to the conduction current. 

The electric field across the third device part, the Nb$_{\text{x}}$O$_{\text{y}}$/Au Schottky contact, depends mainly on the oxygen ion concentration close to the Nb$_{\text{x}}$O$_{\text{y}}$/Au interface. The ions influence the height of the Schottky barrier and the ideality factor. To calculate the electric field across the Schottky contact, the thermionic emission theory is employed. The Schottky contact is described by a set of analytical formulas.\cite{Sze:2007,Tung:2014} Within this theory the Schottky diode current density is given by 
\begin{align}
J_{Schottky} = J_R \left( \exp\{\frac{qV_{Schottky}}{n k_B T}\} - 1 \right),
\label{SchottkyCurrent}
\end{align}
where $k_B$ and $T$ are the Boltzmann constant and temperature, respectively. $n$ is the ideality factor which describes the derivation from an ideal current. The reverse current $J_R$ at forward bias voltages is given by 
\begin{align}
J_R = A^* T^2 \exp\{\frac{-\Phi_b}{k_BT}\},
\label{Reverse1}
\end{align}
where $\Phi_b$ is the Schottky barrier height and $A^* = $ 1.20173 $\times$ 10$^6$ A/(m$^2$K$^2$) the effective Richardson constant. The reverse current in reset direction is dominated by the lowering of the Schottky barrier. If the apparent barrier height $\Phi_b$ at the Schottky interface is significantly smaller than the conductive band gap of the insulator, the reverse current decreases gradually with the applied negative bias approximately as
\begin{align}
J_{R,V < 0} = A^* T^2 \exp\{\frac{-\Phi_b}{k_BT}\}\exp\{\frac{\alpha_r \sqrt{|V|}}{k_BT}\}.
\label{Reverse2}
\end{align}
$\alpha_r$ denotes a device dependent parameter which is used to describe the experimentally observed voltage dependence of the reverse current. Regarding \eqref{SchottkyCurrent}-\eqref{Reverse2} the effective current density is affected by both the ideality factor of the Schottky barrier $n$ and the height of the Schottky barrier $\Phi_b$. The change of the ideality factor is calculated by taking the average distance of the oxygen ions to the Schottky barrier into account,
\begin{align}
n = n_0 +\lambda_n \cdot \bar{d}.
\label{IdealityFactor}
\end{align}
$\lambda_n$ is a fit parameter and $\bar{d}$ is the average relative change of the oxygen ion position from their initial position. $n_0$ is the initial ideality factor which has been estimated to be 4.1 for the device.\cite{Hansen:2015} The change of the Schottky barrier height is attributed to the spatial rearrangement of the mobile oxygen ions during external voltage applications, which leads to an image force adjustment. Taking this into account, the effective Schottky barrier height is calculated by 
\begin{align}
\Phi_b = \Phi_{b0} + \Phi_M,
\label{SchottkyHeight}    
\end{align}
with $\Phi_{b0}$ being the initial Schottky barrier height and $\Phi_M$ being the average surface potential at the Au electrode. This involves in particular the Coulomb field of the ions within the solid state electrolyte which is superimposed to the local electric field produced by the external voltage. Therefore, for a complete description of the electric field the Coulomb field of the oxygen ions has to be taken into account. The Coulomb potential is calculated using Poisson's equation 
\begin{align}
\nabla \cdot \left( \epsilon(\vec{r}) \nabla \Phi(\vec{r}) \right) = -\rho(\vec{r})
\end{align}
where the permittivity $\epsilon(\vec{r}$) is either $\epsilon_{r,\text{Al}_2\text{O}_3}$ or $\epsilon_{r,\text{Nb}_{\text{x}}\text{O}_{\text{y}}}$, the relative permittivity of the Al$_2$O$_3$ or the Nb$_{\text{x}}$O$_{\text{y}}$, respectively.

\section{Results and discussion}

\begin{figure}[t!]
\centering
\includegraphics[width = 8cm]{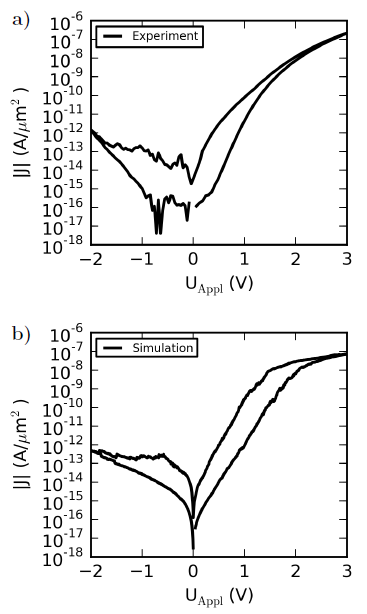}
\caption{a) Measured current-voltage characteristics of the memristive double barrier device. b) Calculated current-voltage characteristics of the memristive double barrier device.}
\label{Kennlinie}
\end{figure}

In figure \ref{Kennlinie} the calculated current-voltage characteristics (also referred to as I-V curve) is compared with a measured I-V curve. For both I-V curves the externally applied voltage is first ramped from 0 to 3 V using a sweep rate of 0.14 V/s in order to set the device resistance from the initial high resistance state (HRS) to a low resistance state (LRS). When the maximum voltage is reached, the sweep rate is switched to -0.14 V/s until the applied voltage is zero again. Then a voltage sweep rate of -0.1 V/s is applied until -2 V is reached. The device is reseted back to the high resistance state. Now the voltage is increased with a sweep rate of 0.1 V/s back to 0 V. As it can be seen from figure \ref{Kennlinie} the calculated I-V curve is in a very good agreement with the experimentally obtained I-V curve. In line with the experimental data the most apparent feature of the memristive hysteresis is the asymmetry between positive and negative bias. This asymmetry can be attributed to the Nb$_{\text{x}}$O$_{\text{y}}$/Au Schottky contact. Furthermore, we are able to capture the nonlinear behavior of the I-V curve in which a high resistance at small voltages and a current saturation at higher voltages is obtained. The model shows moreover the gradual change of the device resistance by several orders of magnitudes. The fine structure for small measured currents at a negative bias which is due to the limitation of the current resolution of the experimental set-up rather due to physically relevant mechanism is not captured by the model.

\begin{figure}[t!]
\centering
\includegraphics[width=13cm]{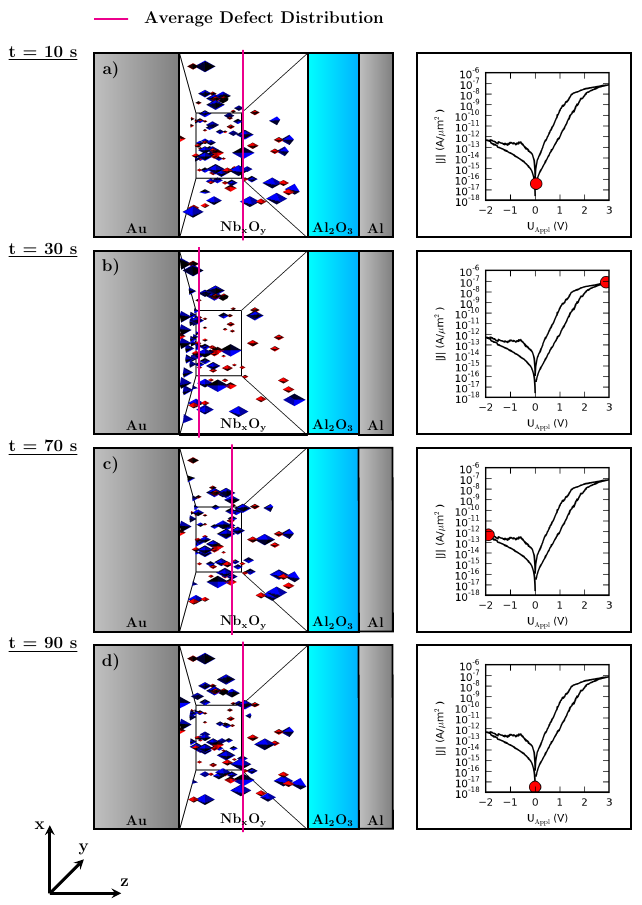}
\caption{The positions of the immobile positively (red) and mobile negatively (blue) charged point defects as an outcome of the simulation are given on the left hand side for four different instants of time. The corresponding position on the current-voltage characteristics is indicated  by a red dot on the right hand side. The colored line represents the average position of the negatively charged point defects at the presented time step.}
\label{3D}
\end{figure}

To study the dynamics of oxygen ions within the solid state electrolyte under the influence of an external voltage and their impact on the device resistance the spatial ion distribution at four significant positions of the I-V curve are shown in a 3D view graph depicted in figure \ref{3D}. At the initial state (figure \ref{3D}a)) the negative and positive ions (blue and red markers, resp.) are concentrated in the center of the simulation box (marked by a vertical line representing the average position of the negative charged ions). By applying an external positive voltage at the left Au electrode the negative charged oxygen ions drift towards the Au interface (figure \ref{3D}b)). When the mobile ions reach the Au interface they can adsorb with a certain probability, which prevents back diffusion. Their average position is then located at the Au interface (see vertical line of figure 3b)). In order to transport oxygen ions back from the Au interface a negative voltage is applied. It is shown in figure \ref{3D}c) that for an external voltage of -2 V the average negative charge position is close to the initial position, while for a complete resetting the external voltage has to be ramped back from -2V to 0V (figure \ref{3D}d)).

\begin{figure}[t!]
\centering
\includegraphics[width = 9cm]{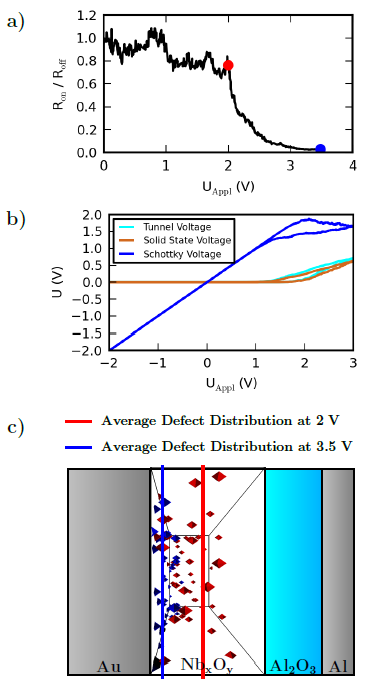}
\caption{a) Resistance ratio between the HRS (R$_\text{off}$) and the LRS (R$_\text{on}$) over the applied voltage for a voltage ramp of 0.14 V/s. b) Voltage drop across the solid state electrolyte, the Schottky barrier, and the tunnel barrier, respectively. c) Calculated position of the negatively charged point defects for an applied voltage of 2 V and 3.5 V.}
\label{IonicInfluence}
\end{figure}

The numerical simulation indicates that the oxygen ion movement is strongly voltage dependent and therefore responsible for the gradual change of the device resistance. To study the voltage dependence of the resistance variation in some more detail, the change of the device resistance as a function of the applied voltage is depicted in figure \ref{IonicInfluence}a). For the data presented in figure \ref{IonicInfluence}a) the sweep rate is set to 0.14 V/s, while the device resistance in the LRS and HRS is determined at a voltage of 0.5 V. If the external voltage is ramped to voltages below 2 V almost no change of the device resistance can be observed. In contrast, if the external voltage is ramped up to 3.5 V the change of the resistance is of several orders of magnitude. Regarding the positions of the negatively charged mobile oxygen ions, as depicted in figure \ref{IonicInfluence}c) as 3D view graph, it can be clearly observed that they only drift in the case of a sufficiently high voltage. In particular, while for a voltage ramp up to 2 V the ion positions are nearly unaffected compared to the initial distribution. For a voltage ramp up to 3.5 V nearly all negative charged ions are located at the Au interface. 

To get an idea of the physical mechanisms behind the strong voltage dependence, the voltage drops across the different device parts are given in figure \ref{IonicInfluence}b). These are the voltages across the tunnel barrier, the Nb$_{\text{x}}$O$_{\text{y}}$ solid state electrolyte, and the Schottky contact. At low applied voltages the partial voltages across the Nb$_{\text{x}}$O$_{\text{y}}$ layer and the tunnel layer (orange and green curves of figure \ref{IonicInfluence}b)) are almost zero since the current is blocked by the Schottky contact (blue curve of fig. \ref{IonicInfluence}b)). In other words, the Schottky contact defines a threshold voltage for the device which has to be exceeded in to change the resistance of the device. Here, the resulting electric field within the Nb$_{\text{x}}$O$_{\text{y}}$ is too small to induce ionic diffusion. However, since the externally applied voltage exceeds 1.5 V, the voltage across the Nb$_{\text{x}}$O$_{\text{y}}$ layer raises and excites voltage polarity directed ion transport. Moreover, taking figure \ref{IonicInfluence}b) into account, the reset process is mainly dominated by the voltage drop across the Schottky barrier. Due to the reverse direction of the diode like Schottky contact nearly the complete applied voltage is blocked by the Schottky contact during the reset process. The increasing voltage drop across the Schottky contact is able initiate desorption of oxygen ions from the Au surface (E$_{ij} = 0.25$ eV \cite{Gross:2004}). Then the desorbed ions drift subject to the Coulomb potential of the remaining positive charges and subject to the concentration gradient back into the solid state electrolyte. This mechanism leads to a fast reset of the total device resistance. 
    
\begin{figure}[t!]
\center
\includegraphics[width=12cm]{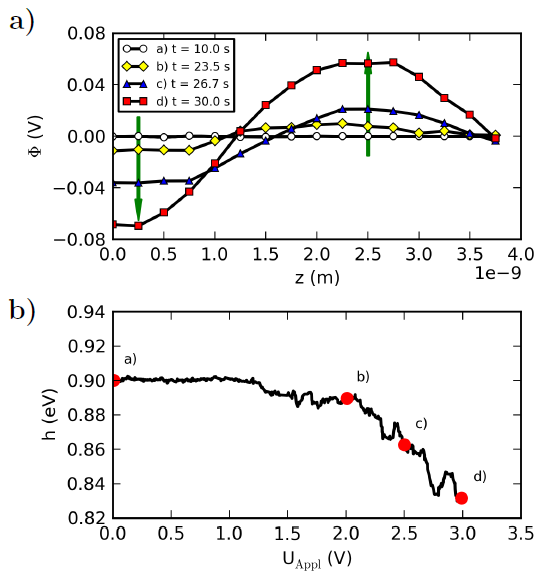}
\caption{a) The in $x$ and $y$ directions averaged Coulomb potential due to the charged defects, plotted as a function of $z$ is shown for different time steps. The transport of negative charged point defects within the field of a positive applied voltage at the Au electrode leads to a charge separation and therefore to a positive Coulomb-potential within the device, and thus to a negative Coulomb-potential at the Schottky contact ($z = 0$). The corresponding change of the Schottky barrier height due to the negative Coulomb potential plotted as a function of the applied voltage is shown in b). }
\label{SchottkyBarrierHeight}
\end{figure}
 
The reason for the drastic resistance change of the device stems hypothetically from the variation of the two energy barriers which confine the solid state electrolyte.\cite{Hansen:2015} These variations are induced by the change of the spatial oxygen concentration, as discussed above. This leads to a kind of interference of the both interfaces which embed the Nb$_{\text{x}}$O$_{\text{y}}$ layer. In line with \eqref{Tunnellength}, \eqref{IdealityFactor}, and \eqref{SchottkyHeight} we assume for the model that the spatial distribution of negative charged oxygen ions in front of the barriers influences both the effective tunnel barrier length and the parameters of the Schottky barrier, such as the ideality factor and the barrier height. We find that the effective tunnel barrier width decreases from 1.3 nm to 1.2 nm if the voltage applied to the Au electrode raises from 0 V to 3.0 V. At the same time the ideality factor decreases from 4.0 to 3.4. The Schottky barrier height is lowered from 0.9 eV to 0.83 eV (see figure \ref{SchottkyBarrierHeight}b)). 

As given in \eqref{Tunnellength} and \eqref{IdealityFactor} both the effective tunnel distance and the ideality factor are assumed to depend linearly on the average ion position, while the lowering of the Schottky barrier height depends on the Coulomb potential in front of the Au interface (see  \eqref{SchottkyHeight}). The variation of the Coulomb potential as a function of the applied external voltage is shown in figure \ref{SchottkyBarrierHeight}a). For this study the local Coulomb potential is averaged along the $x$ and $y$ directions. We find in particular that the Coulomb potential at the Nb$_{\text{x}}$O$_{\text{y}}$/Au interface decreases from 0 mV to -70 mV when ramping the external voltage from 0 V to 3.0 V. This means that the lowering of the Schottky barrier height (see figure \ref{SchottkyBarrierHeight}b)) originates from the displacement of negatively charged ions towards the Au interface, which decreases the Coulomb potential at the interface. 

\begin{figure}[t!]
\centering
\includegraphics[width =12cm]{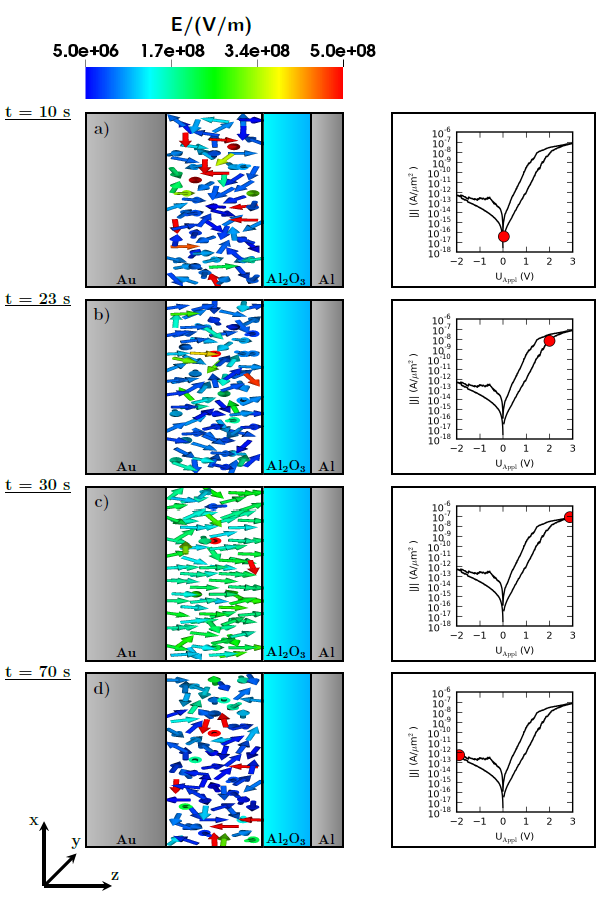}
\caption{Projection of the local electric field onto the $xz$-plane for distinct instants of time.}
\label{ElectricField}
\end{figure}

\begin{figure}[t!]
\centering
\includegraphics[width = 8cm]{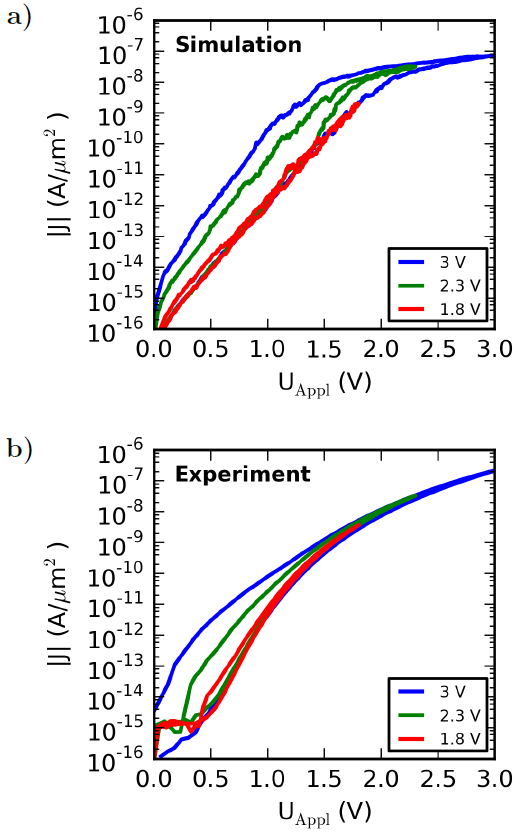}
\caption{I-V curves for three different maximum applied voltages (1.8 V, 2.3 V and 3 V). a) simulation results, b) experimental results. 
}
\label{SimulationMeasurement}
\end{figure}

In order to identify the rate dependent mechanisms of the ion dynamics, the electric field within the Nb$_{\text{x}}$O$_{\text{y}}$ layer is depicted in figure \ref{ElectricField} at four distinct instances of time. To visualize the actual mechanism the electric field is projected onto the $xz$-plane. For $t = 10$ s the applied voltage is 0 V and the origin of the electric field are the positive and negative charges. This attractive Coulomb field has to be overcome to deflect charges from the stable position. Small setting voltages are blocked by the Schottky barrier (see figure \ref{IonicInfluence}a)). However, for setting voltages larger than approx. 1.5 V the voltage drop across the solid state electrolyte starts to increase. For an applied voltage of 2 V a small preferred orientation of the electric field in positive $z$ direction can be observed (see figure \ref{ElectricField}b)) although the random structure of the Coulomb field can still be seen. This preferred orientation and the strength of the electric field increases for higher applied voltages up to 3 V (see figure \ref{ElectricField}c)). A positive voltage at the Au electrode on the device leads to a homogenization of the electric field. Although between 2 V and 3 V at certain positions the Coulomb field is still dominant and prevents ionic motion. The electric field due to the applied voltage is dominant so that an ionic drift occurs. 

These observations indicate that the maximum applied voltage has a strong influence on the shape of the I-V curve in set direction. In particular, since for voltages smaller than approx. 1.5 V the directed ionic motion is suppressed. Furthermore, the hysteresis of the I-V curve is expected to disappear for maximum applied voltages smaller than approx. 1.5 V. To verify this hypothesis the positive branch of simulated and measured I-V curves for three different maximum applied voltages (1.8 V, 2.3 V and 3 V) are shown in figure \ref{SimulationMeasurement}. The simulated and measured I-V curves are obtained under comparable conditions. Both I-V curves show a nearly vanishing hysteresis for a maximum applied voltage of 1.8 V . However, for higher maximum applied voltages the shape of the hysteresis becomes significantly broader. These findings can be explained by ionic motion rather than by charging and de-charging of trap states. Therefore our results support ionic motion as a key feature for resistance change within the memristive double barrier device.

\begin{figure}[t!]
\center
\includegraphics[width = 12cm]{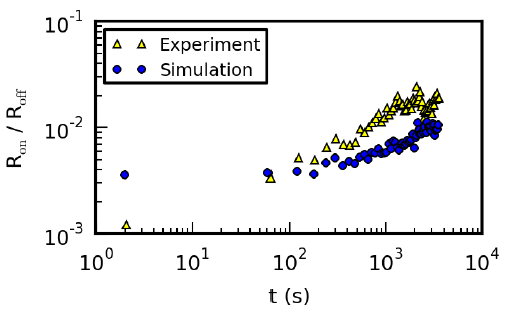}
\caption{Retention characteristics of the memristive device. The time development of the LRS related to the HRS is plotted as a function of time. The yellow triangles represent the measured and the blue circles the simulated retention characteristics. For the readout of the LRS pulses of 0.5 V are applied every 60 s in the experiment as well as in the simulation.}
\label{Adsorption}
\end{figure}

In order to investigate the retention characteristics of the device, it is set to the LRS which is the initial state for this study. The change of the device resistance is then mapped over time. To track the evolution the resistance of the device, voltage pulses of 0.5 V are applied every 60 s in order to readout the resistance. Figure \ref{Adsorption} shows the simulation results as well as experimental results for the same procedure. The results show a long resistance retention up to days without an externally applied voltage. Since almost the complete voltage during the reset process is blocked by the Schottky contact, the diffusion of the negative oxygen ions is similarly with or without an applied reset voltage. However, the desorption mechanism strongly depends on the Schottky contact and therefore on the applied reset voltage. The significantly different time scales of the change of the device resistance with or without an applied reset voltage indicate that the desorption mechanism is in fact the time limiting process of the resistance retention rather than the back diffusion of free oxygen ions into the bulk of the device. A fast resistance change during the first 60 s (see figure \ref{Adsorption}) which can be found in the experimental results but not in the simulation results is explained by the fast de-charging of trap states at the Nb$_{\text{x}}$O$_{\text{y}}$/Au interface. This mechanism is actually not included in the simulation.\cite{Hansen:2015} Most probably both mechanisms, charge trapping and ion drift are present in the devices but leading to retention times on different time scales. It is important to note that the local electric field and the resulting Coulomb potentials have to be analyzed in order to determine the rate dependent mechanisms of the ion dynamics within the numerical investigations.

\section{Conclusions}

We report on numerical simulations of a quantum mechanical double barrier memristive device. The device consists of an Al/Al$_2$O$_3$/Nb$_{\text{x}}$O$_{\text{y}}$/Au structure where the Nb$_{\text{x}}$O$_{\text{y}}$ solid state electrolyte is enclosed by a tunnel barrier and a Schottky contact. We present results from a kinetic Monte Carlo simulation of the (atomistic) ion transport, consistently coupled to a lumped circuit element model for the current through the device. The current voltage characteristics which is an important indicator for the dynamics of the device is numerically calculated and compared to experimentally obtained results. We find very good agreement between simulation and experiment. We claim that the model which is proposed captures most relevant physical processes for resistive switching on both, atomistic length scales and experimental time scales. The simulation results identify the transport of charged point defects within the solid state electrolyte as a key mechanism for the resistive switching of the double barrier device, although the charging and de-charging of trap states within the solid state electrolyte and/or at the metal semiconductor interface cannot be completely excluded from the physical picture. The results indicate that the local electric field is the rate determining quantity, particularly during the set process. Finally, we find that the threshold voltage for resistive switching in set direction and an absorption mechanism is main reason for the long retention time. It is found that the Coulomb field and the concentration gradient are the main reasons for back diffusion of charged point defects. The calculated retention time is in excellent agreement with experimental findings. We believe that with the model we provide a powerful simulation tool that can be used to gain a deeper physical understanding of the resistive switching mechanisms of the double barrier memristive device.

\section*{Acknowledgement}

The authors gratefully acknowledge financial support provided by the Deutsche Forschungsgemeinschaft DFG in the frame of Research Group For 2093 ``Memristive Devices for Neural Systems''.

\end{document}